\documentclass[aps,prl,reprint,groupedaddress,floatfix]{revtex4-1}
\usepackage[utf8]{inputenc}

\usepackage{amssymb,amsmath,bm,bbold,graphicx}
\usepackage{ifpdf,color,mathrsfs}

\definecolor{darkblue}{rgb}{0,0,.6}
\definecolor{darkgreen}{rgb}{0,0.5,0}

\ifpdf
\usepackage{epstopdf}
\usepackage[pdftex,unicode,
pdfstartview={FitH},pdfborder={0 0 0}]{hyperref}
\usepackage{hypcap}
\else
\usepackage[hypertex]{hyperref}
\fi
\hypersetup{
bookmarksnumbered = true,
colorlinks = true, linkcolor = darkblue,
citecolor = darkblue, filecolor = darkblue,
menucolor = darkblue, urlcolor = darkblue
}

\newcommand{\iu}{i} 
\newcommand{\de}{d} 
\newcommand{\ee}{e} 
\newcommand{\abs}[1]{\ensuremath{\left| #1 \right|}}

\let\Im\undefined

\DeclareMathOperator{\Im}{Im}

\newcommand{\ket}[1]{\ensuremath{\left| #1\right\rangle}}

\newcommand{\melement}[3]{\ensuremath{\left\langle #1\left| #2 \right| #3\right\rangle}}

\begin{document}

\title{Transient optical gain in strong-field-excited solids}

\author{Muhammad Qasim}
\affiliation{Max-Planck-Institut f\"ur Quantenoptik, Hans-Kopfermann-Str. 1, Garching 85748, Germany}
\affiliation{Ludwig-Maximilians-Universit\"at, Am Coulombwall~1, Garching 85748, Germany}

\author{Dmitry A.~Zimin}
\affiliation{Max-Planck-Institut f\"ur Quantenoptik, Hans-Kopfermann-Str. 1, Garching 85748, Germany}
\affiliation{Ludwig-Maximilians-Universit\"at, Am Coulombwall~1, Garching 85748, Germany}


\author{Vladislav S.~Yakovlev}
\email[]{vladislav.yakovlev@mpq.mpg.de}
\affiliation{Max-Planck-Institut f\"ur Quantenoptik, Hans-Kopfermann-Str. 1, Garching 85748, Germany}
\affiliation{Ludwig-Maximilians-Universit\"at, Am Coulombwall~1, Garching 85748, Germany}

\date{\today}

\begin{abstract}
	Multiphoton excitation of a solid by a few-cycle, intense laser pulse forms a very non-equilibrium distribution of charge carriers, where occupation probabilities do not necessarily decrease with energy.
	We show that, under certain conditions, significant population inversion can emerge between pairs of valence- or conduction-band states, where transitions between the Bloch states are dipole-allowed.
	This population inversion leads to stimulated emission in a laser-excited solid at frequencies where the unperturbed solid is transparent.
	We establish the optimal conditions for observing the strong-field-induced optical gain.
\end{abstract}


\maketitle

Ultrafast photoinjection of charge carriers is one of the basic approaches for the control of optical properties of solids on ultrashort time scales.
The relevant effects include the well-studied Drude-like polarization response of photoinjected carriers~\cite{Gamaly_PQE_2013}, renormalization of band energies~\cite{Spataru_PRB_2004, Haug-Koch_2009, Schultze_Science_2014}, disappearance of excitonic resonances~\cite{Wake_PRB_1992, Huber_PRB_2005}, changes in the nonlinear polarization response of a solid~\cite{Boyd_2008}, and also the formation of transient population inversion, the origin of which is different from that described in basic textbooks on laser physics.
For example, population inversion and the associated optical gain appear in optically pumped graphene due to relaxation bottleneck~\cite{Gierz_JPCM_2015}.
Population inversion was also observed in atomically thin WS${}_2$, where the main physical mechanism was found to be the bandgap renormalization in the presence of excitons~\cite{Chernikov_NP_2015}: once the band edge is reduced below the energy of the exciton resonance, the decay of an exciton may release energy in the form of optical gain.
Excitons were reported to be responsible for transient gain in photoexcited quantum wells~\cite{Lange_PRB_2009}.
Also, two-photon stimulated emission was observed in sapphire excited by multiphoton absorption~\cite{Winkler_NP_2018}.

We investigate yet another type of transient population inversion, which is characteristic to transparent solids nonlinearly photoexcited by a few-cycle laser pulse, the central frequency of which is much smaller than the bandgap.
When an electron is excited from a valence state to a conduction state by multiphoton absorption or interband tunneling, the transition probability is known to be very sensitive to the energy difference between the states: the larger the energy gap, the smaller the excitation probability.
Nevertheless, these are not always states in the uppermost valence band that are depleted most.
When the laser pulse depletes one of the deeper valence states more efficiently than a state above it, population inversion emerges.
A similar effect is well-known for molecules: strong-field ionization does not always favor the highest occupied molecular orbital~\cite{Smirnova_Nature_2009}.
If this kind of population inversion occurs in bulk solids, it may be possible to observe it as optical gain in the spectral region where the unperturbed crystal is transparent; however, fast relaxation processes make this gain short-lived.
Once photoinjected charge carriers thermalize, which usually takes a few tens of femtoseconds~\cite{Shah_1999, Rohde_PRL_2018}, no population inversion between valence-band states is possible.
Therefore, it is important to clarify the optimal conditions for the observation of such anomalies in the depletion of valence-band states.
This is the main purpose of this Letter.

Let us consider two electrons that share the same crystal momentum $\mathbf{k}$ and initially reside in fully occupied valence bands $v_1$ and $v_2$.
A laser pulse depletes the valence-band states by promoting the electrons to conduction bands.
Can a laser pulse deplete the energetically lower $v_1$ state more efficiently than the upper $v_2$ state, thus forming population inversion?
Usually, this does not happen because the probability of multiphoton or tunneling excitation rapidly decreases with the transition energy.
However, this probability also depends on the dipole transition matrix element between the initial valence- and final conduction-band states.
Let us consider this dependence in the case of a homogeneous electric field interacting with an electron in a periodic lattice potential.
In the basis of accelerated Bloch states~\cite{Krieger_1986_PRB}, the mathematical expression that is responsible for exciting an electron from a valence band $v$ to the lowest conduction band $c$ at a time $t$ contains the scalar product of the laser field and the dipole transition matrix element: $\mathbf{F}_L(t)\cdot \mathbf{d}_{c v}\bigl(\mathbf{k} +e\hbar^{-1} \mathbf{A}_L(t)\bigr)$.
Here, $e>0$ is the elementary charge and $\mathbf{F}_L(t) = -\mathbf{A}_L'(t)$ is the external electric field.
It is important to note that the magnitude of $\mathbf{d}_{c v}$ does not need to be small to suppress excitations from the upper valence state---if the vectors $\mathbf{F}_L(t)$ and $\mathbf{d}_{c v}\bigl(\mathbf{k} +e\hbar^{-1} \mathbf{A}_L(t)\bigr)$ were orthogonal to each other at all times, the laser pulse would not cause direct transitions from band $v$ to band $c$ at crystal momentum $\mathbf{k}$.
This suggests the following recipe for achieving population inversion between two valence-band states: among all the directions orthogonal to $\mathbf{d}_{c v_2}(\mathbf{k})$, choose the one that is most aligned with $\mathbf{d}_{c v_1}(\mathbf{k})$ and use $\mathbf{F}_L$ polarized along this direction.
There are, however, many reasons why this simple recipe may not work.
The above analysis neglects transitions from $v_2$ to upper conduction bands; it also neglects transitions among conduction bands and among partially depleted valence-band states.
Also, for the full suppression of direct transitions from $v_2$ to $c$, the condition $\mathbf{F}_L \cdot \mathbf{d}_{c v_2} = 0$ must be satisfied not just for a particular crystal momentum $\mathbf{k}$, but along the line $\mathbf{k} +e\hbar^{-1} \mathbf{A}_L(t)$.
Finally, the components of $\mathbf{d}_{c v}$ are, in general, complex-valued, so that no real-valued vector $\mathbf{F}_L$ that is perpendicular to $\mathbf{d}_{c v_2}$ may exist.
In the rest of this Letter, we numerically demonstrate that, in spite of all these complications, the orientation of the $\mathbf{d}_{c v}$ vectors plays a crucial role in forming a population inversion and the associated optical gain.

\begin{figure}
	\includegraphics[trim={1mm 1mm 0 0},clip,width=0.95\columnwidth]{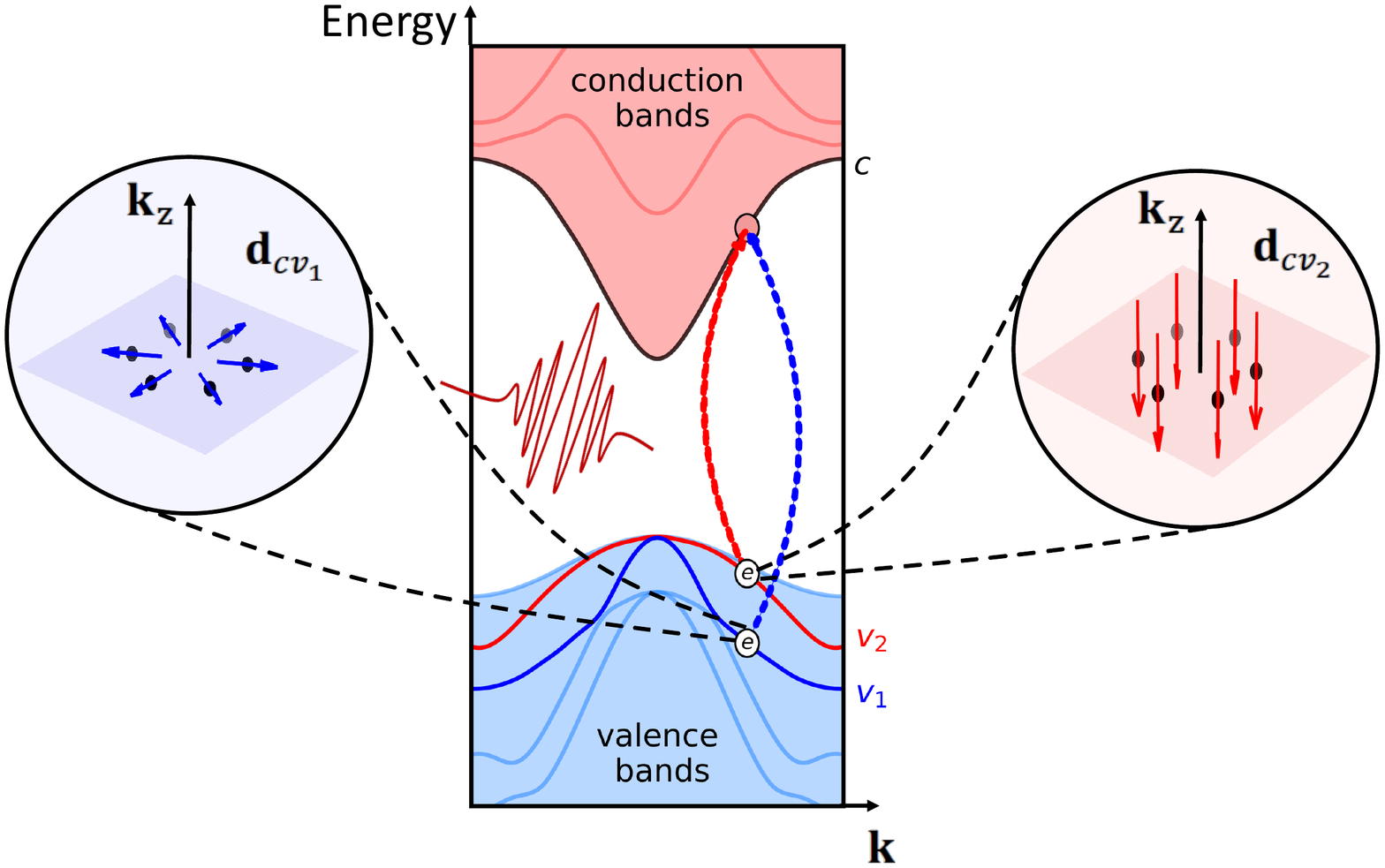}
	\vspace{-8mm}
	\caption{\label{fig:the_concept}
		The ideal setting for the formation of population inversion between valence bands $v_1$ and $v_2$ at crystal momenta that form a star of $\mathbf{k}$.
		The left and right insets show the $\mathbf{d}_{c v_1}\left(\mathbf{k}\right)$ and $\mathbf{d}_{c v_2}\left(\mathbf{k}\right)$ vectors, respectively.
		These matrix elements control the probabilities of photoexcitation to the lowest conduction band.
		If the electric field of the photoexciting laser pulse is perpendicular to the vertical axis, then $\mathbf{F}_L \cdot \mathbf{d}_{c v_2}(\mathbf{k}) \equiv 0$, so that all the $\{v_2,\mathbf{k}\} \to \{c,\mathbf{k}\}$ transitions are strongly suppressed.
		At the same time, the laser pulse will efficiently drive some of the $\{v_1,\mathbf{k}\} \to \{c,\mathbf{k}\}$ transitions.
	}
\end{figure}

The presence of a population inversion between two valence-band states is not sufficient for observing optical gain, even if the transition between the states is dipole-allowed.
This is because the electronic structure of a solid supports, in general, multiple single-photon transitions at a given frequency.
Stimulated emission due to one of these transitions can be compensated by absorption due to another one that lacks population inversion.
To establish the optimal conditions for observing optical gain, we must consider the crystal symmetry.
By applying all the point-group symmetry operations $R$ to a crystal momentum $\mathbf{k}$, we obtain a star of $\mathbf{k}$ \cite{Dresselhaus_2007}: a set of crystal momenta with a shared set of band energies $\epsilon_n(\mathbf{k}) = \epsilon_n(R \mathbf{k})$ and dipole matrix elements that are related to each other by $\mathbf{d}_{n m}(R \mathbf{k}) = R \mathbf{d}_{n m}(\mathbf{k})$.
Our goal is to suppress transitions from band $v_2$ in a particular star of $\mathbf{k}$ knowing that $\mathbf{F}_L \cdot \mathbf{d}_{c v_2}(\mathbf{k}) = 0$ for one of its elements.
This requirement translates into
\begin{equation}
	\label{eq:star_orthogonality}
	0 = \mathbf{F}_L \cdot \mathbf{d}_{c v_2}(R \mathbf{k}) = \mathbf{F}_L \cdot R \mathbf{d}_{c v_2}(\mathbf{k})\ \text{for all $R$}.
\end{equation}
The point-group symmetry operations consist of rotations and, possibly, inversion.
To satisfy the above condition, all the rotations must be around the same axis. 
If $\mathbf{d}_{c v_2}(\mathbf{k})$ is parallel to the rotation axis, then all the $\mathbf{d}_{c v_2}(R \mathbf{k})$ vectors are parallel to each other, and Eq.~\eqref{eq:star_orthogonality} is satisfied whenever $\mathbf{F}_L$ is orthogonal to the rotation axis.
This is the case illustrated in the right inset of Fig.~\ref{fig:the_concept}.
Alternatively, Eq.~\eqref{eq:star_orthogonality} can be satisfied if $\mathbf{F}_L$ is parallel to the rotation axis, while all the $\mathbf{d}_{c v_2}(R \mathbf{k})$ vectors are orthogonal to it.
Even though Eq.~\eqref{eq:star_orthogonality} does not guarantee the formation of population inversion in the entire star of $\mathbf{k}$, we find it to be an excellent indicator for observing optical gain in a nonlinearly excited solid.

For this purpose, crystals that possess only one axis of rotation are particularly well suited.
Also, we need the laser pulse to preserve its polarization state as it propagates through the medium; otherwise, Eq.~\eqref{eq:star_orthogonality} may not hold throughout the entire sample.
The polarization is preserved if the laser beam propagates along the optic axis of a uniaxial crystal, unless the pulse induces a strong birefringence by exciting the crystal~\cite{Qasim_PRB_2018}.
So we choose $\mathbf{F}_L$ to be perpendicular to the crystal axis.
Searching for conditions that favor the appearance of stimulated emission, we are looking for pairs of valence-band states where $\mathbf{d}_{c v_2}(\mathbf{k})$ is parallel to the crystal axis, while $\mathbf{d}_{c v_1}(\mathbf{k})$ is perpendicular to it, as shown in Fig.~\ref{fig:the_concept}.

In the following, we illustrate the above ideas using gallium nitride (GaN) as a representative uniaxial crystal.
We model its interaction with light by solving the velocity-gauge time-dependent Schr\"odinger equation (TDSE)
\begin{equation} \label{eq:EOM}
\iu \hbar \dfrac{\de}{\de t} \ket{\psi_{n \mathbf{k}}(t)} =\left( \hat{H}_{\mathbf{k}}^{(0)} + \frac{e}{m_0}\mathbf{A}_L(t) \cdot \hat{\mathbf{p}} \right) \ket{\psi_{n \mathbf{k}}(t)}
\end{equation}
for a set of initial valence-band states.
Here, $m_0$ is the free-electron mass and $\hat{\mathbf{p}}$ is the momentum operator.
We expand the wavefunctions in the basis of stationary three-dimensional Bloch states:
\begin{equation}\label{eq:psi_ansatz}
\ket{\psi_{n \mathbf{k}}(t)} = \sum_m \alpha_{m n}(\mathbf{k}, t)
\ee^{-\frac{\iu}{\hbar} \epsilon_{m}(\mathbf{k}) t}
\ket{m \mathbf{k}}.
\end{equation}
The energies, $\epsilon_{m}(\mathbf{k})$, and the eigenstates of the unperturbed Hamiltonian are  defined by $\hat{H}_{\mathbf{k}}^{(0)} \ket{m \mathbf{k}} = \epsilon_{m}(\mathbf{k}) \ket{m \mathbf{k}}$. 
An expansion coefficient $ \alpha_{mn} (\mathbf{k}, t) $ is equal to the probability amplitude of observing an electron in state $ \ket {m \mathbf{k}} $, provided that the electron was originally in state $\ket {n \mathbf{k}}$.
Consequently, the initial condition for solving Eq.~\eqref{eq:psi_ansatz} is $ \alpha_{mn} (\mathbf{k}, t_0) = \delta_{mn} $ with $n$ being a valence band.

In this formalism, the electronic structure of the solid is fully described by its band energies  $\epsilon_{m}(\mathbf{k})$ and transition matrix elements $\mathbf{p}_{m n}(\mathbf{k}) = \melement{m \mathbf{k}}{\hat{\mathbf{p}}}{n \mathbf{k}}$, which we obtain from density functional theory (DFT) with the Tran-Blaha exchange-correlation potential~\cite{Tran_PRL_2009}, using the Elk code~\cite{ELK_code}.
These matrix elements are related to the dipole transition matrix elements by $\mathbf{d}_{m n} = - \iu e \hbar \mathbf{p}_{m n} / [m_{0} (\epsilon_{m}- \epsilon_{n})] $. 
We discretized reciprocal space by the $20 \times 20 \times 20$ Monkhorst-Pack grid and used a basis of 100 energy bands, 18 of which were valence bands.

We parameterize the laser pulse with
\begin{equation}
\mathbf{A}_L(t) = -\mathbf{e}_L \frac{F_L}{\omega_L} \theta(T_L - |t|)
\cos^4 \left( \frac{\pi t}{2 T_L} \right) \sin(\omega_L t),
\end{equation}
where $\mathbf{e}_L$ is a unit vector along the laser polarization, $\omega_L$ is the central frequency, $\theta(t)$ is the Heaviside function, and $T_L$ is related to the full width at half maximum (FWHM) of the pulse by $T_L = \pi\,\mathrm{FWHM} / \left(4\,\text{arccos}(2^{-0.125})\right)$.
We used $\hbar\omega_L = 1.56$~eV (800~nm), $\mathrm{FWHM} = 4$~fs, and 
$\mathbf{e}_L$ pointing along the $[1 \bar{1} 0]$ direction, which is parallel to the mirror plane of the GaN crystal.

\begin{figure}
	\begin{tabular}{p{0.96\columnwidth} p{0pt}}
		\vspace{0mm} \includegraphics[trim={0 0 1mm 0},clip,width=0.90\columnwidth]{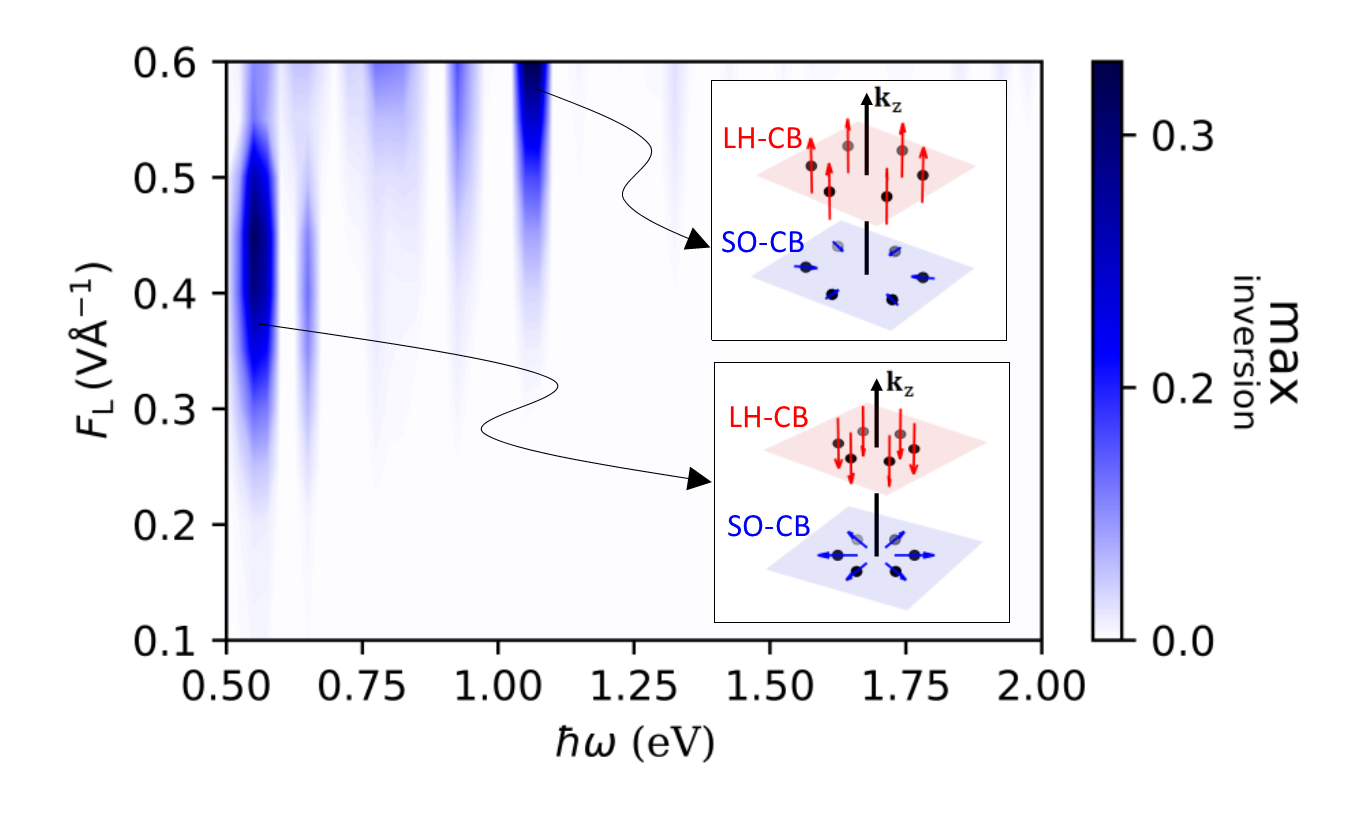} &
		\vspace{-0mm} \hspace{-0.99\columnwidth}
		\textbf{(a)}
	\end{tabular}
	\begin{tabular}{p{0.96\columnwidth} p{0pt}}
		\vspace{-5mm} 
		\includegraphics[width=0.90\columnwidth]{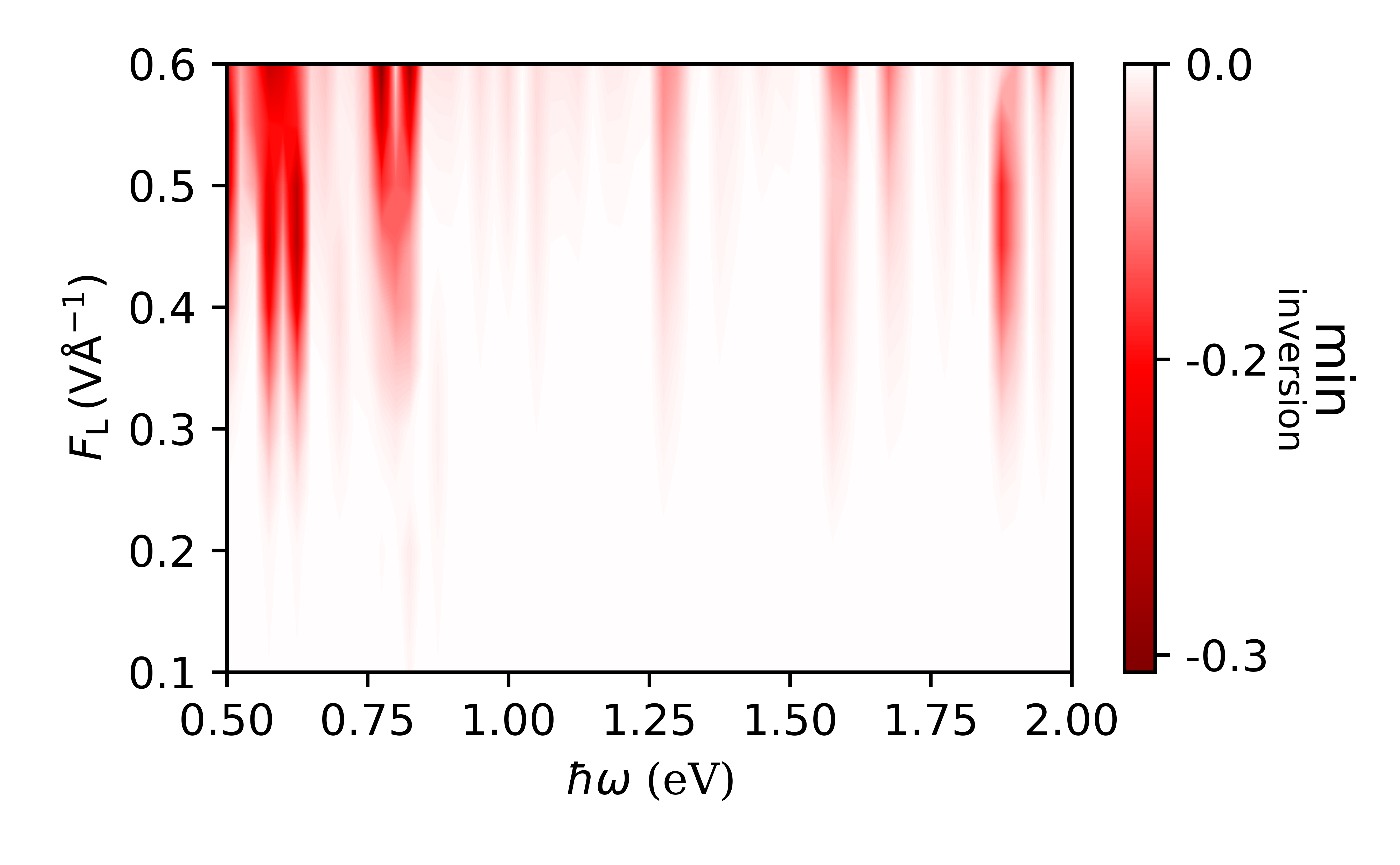}&
		\vspace{-4mm} \hspace{-0.99\columnwidth}
		\textbf{(b)}
	\end{tabular}
	\vspace{-5mm}
	\caption{\label{fig:population_inversion}
		The maximal (a) and minimal (b) population inversion in dependence of transition energy and peak laser field.
		The insets in (a) show the magnitudes and orientations of $\mathbf{d}_{c v}(\mathbf{k})$
		for the two stars of $\mathbf{k}$ where population inversion is particularly prominent;
		LH, SO, and CB denote the light hole, split-off, and conduction bands.
	}
\end{figure}

From the numerical solution of Eq.~\eqref{eq:EOM}, we evaluate the occupation probabilities:
\begin{equation}
f_m(\mathbf{k}) = \sum_{m' \in \text{VB}} |\alpha_{m m'}(\mathbf{k}, T_L)|^2,
\end{equation}
where we add the contributions from all the valence bands (VB).
Population inversion emerges whenever $f_m(\mathbf{k}) > f_n(\mathbf{k})$ for $\epsilon_{m}(\mathbf{k}) > \epsilon_{n}(\mathbf{k})$.
Figure~\ref{fig:population_inversion} displays the maximal and minimal values of $f_m(\mathbf{k}) - f_n(\mathbf{k})$.
More precisely, Fig.~\ref{fig:population_inversion}(a) displays
	$ \max_{\abs{\epsilon_{m}(\mathbf{k})  - \epsilon_{n}(\mathbf{k}) - \epsilon} \lesssim \Delta\epsilon}\{ f_{m}\left(\mathbf{k},F_L\right) - f_{n}(\mathbf{k},F_L)\} $;
we observe significant ($\sim 0.3$) population inversion for $F_L\gtrsim 0.3$~V\AA$^{-1}$.
However,
	$ \min_{\abs{\epsilon_{m}(\mathbf{k})  - \epsilon_{n}(\mathbf{k}) - \epsilon} \lesssim \Delta\epsilon}\{ f_{m}\left(\mathbf{k},F_L\right) - f_{n}(\mathbf{k},F_L)\}$, 
shown in Fig.~\ref{fig:population_inversion}(b), tells us that there are also transitions that contribute to absorption in the same spectral regions where we find transitions that would amplify probe light.
For both figures, we used an energy bin of $\Delta\epsilon = 0.025$~eV.

According to Fig.~\ref{fig:population_inversion}(a), population inversion mainly forms  at $\hbar\omega=0.6$~eV and $\hbar\omega=1.1$~eV, and these frequencies are practically independent of the peak field strength of the laser pulse. 
Analyzing the occupation probabilities, we see that, for both these frequencies, the relevant lower and upper Bloch states reside in the split-off (SO) and light-hole (LH) bands. 
The insets of Fig.~\ref{fig:population_inversion}(a) display the stars of $\mathbf{k}$ associated with these transitions. 
Both these stars lie in the $k_z=0$ plane.
The crystal momenta that are responsible for transitions at 0.6~eV have $|\mathbf{k}|=0.14$~\AA$^{-1}$, while the 1.1-eV transitions are due to crystal momenta with $|\mathbf{k}|=0.21$~\AA$^{-1}$.
The arrows in the insets point along those directions of linearly polarized light, $\mathbf{e}_L$, that maximize $\abs{\mathbf{e}_L \cdot \mathbf{d}_{c v}(\mathbf{k})}$, while the length of each arrow represents $|\mathbf{d}_{c v}(\mathbf{k})|$.
The laser field in these simulations was perpendicular to $\mathbf{k}_{z}\parallel[001]$, so excitations from the LH band were strongly suppressed at both stars of $\mathbf{k}$ (the red arrows are approximately orthogonal to $\mathbf{e}_L$).
In contrast, some of the states in the energetically lower SO band have their $\mathbf{d}_{c v_{1}}(\mathbf{k})$ aligned with the laser field.
These states are depleted significantly, which leads to population inversion.

As long as there are amplifying and absorbing transitions at the same frequencies, the above analysis does not tell us whether a photo-excited solid will amplify or attenuate a weak probe pulse.
To find this out, we calculated the tensor of linear susceptibility~\cite{Qasim_PRB_2018}:
\begin{multline} \label{eq:chi}
\chi_{\alpha \beta}(\omega) =
e^2 \int_{\mathrm{BZ}} \frac{\de^3 \mathbf{k}}{\left(2\pi \right)^3}\, 
\sum_n f_n(\mathbf{k}) \biggl\{-\frac{\hat{m}^{-1}(n, \mathbf{k})}{\omega^2}  +  \frac{1}{\hbar m_0^2} \\
\times
\sum_{m \ne n} \biggl[ \frac{1}{\omega_{m n}^2(\mathbf{k})} \biggl(
\frac{2 \iu}{\omega + \iu \gamma}
\Im\left[ p_{n m}^{\alpha}(\mathbf{k}) p_{m n}^{\beta}(\mathbf{k}) \right] \\+
\frac{p_{n m}^{\alpha}(\mathbf{k}) p_{m n}^{\beta}(\mathbf{k})}
{\omega_{m n}(\mathbf{k}) - \omega - \iu \gamma} +
\frac{p_{m n}^{\alpha}(\mathbf{k}) p_{n m}^{\beta}(\mathbf{k})}
{\omega_{m n}(\mathbf{k}) + \omega + \iu \gamma}
\biggr) \biggr] \biggr\}.
\end{multline}
Here, $\hat{m}^{-1}$ is the inverse-mass tensor, while $\gamma$ is a phenomenological dephasing rate.
Our main motivation for introducing the dephasing parameter was not to account for physical processes that destroy interband coherence, but to counteract numerical artifacts---without dephasing, the absorption spectrum would consist of a set of sharp resonances that correspond to transitions at the nodes of the $\mathbf{k}$ grid.
For this purpose, we chose $\gamma = 5 \times 10^{13}$~s ($\gamma^{-1} = 20$~fs).

\begin{figure}
	\begin{tabular}{p{0.96\columnwidth} p{0pt}}
		\vspace{0mm}
		 \includegraphics[trim={0 0 1mm 0},clip,width=0.96\columnwidth]{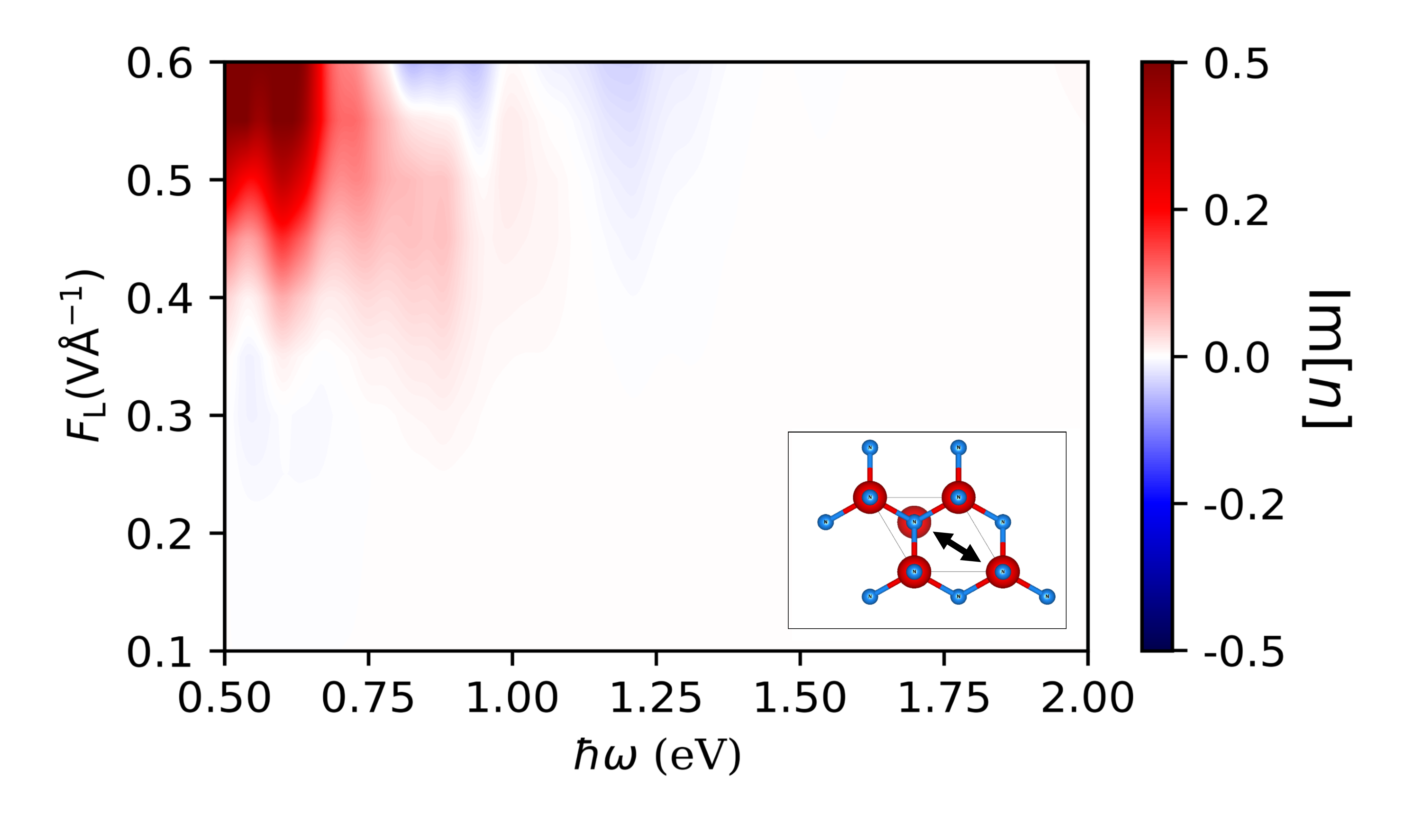} &
		\vspace{-0mm} \hspace{-0.99\columnwidth}
		\textbf{(a)}
	\end{tabular}
	\begin{tabular}{p{0.96\columnwidth} p{0pt}}
		\vspace{-5mm} \includegraphics[trim={0 0 1mm 0},clip,width=0.96\columnwidth]{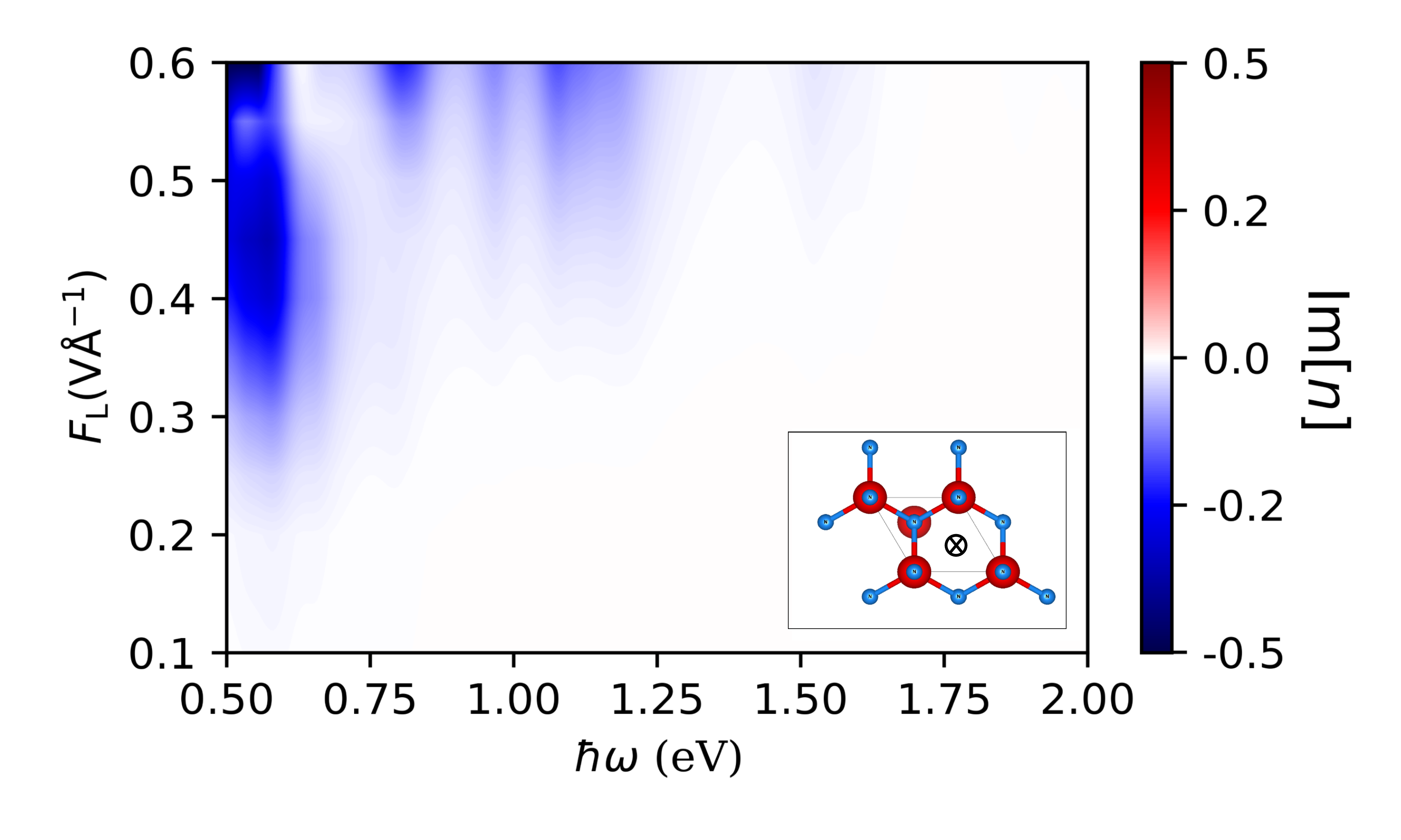}&
		\vspace{-4mm} \hspace{-0.99\columnwidth}
		\textbf{(b)}
	\end{tabular}
	\vspace{-5mm}
	\caption{\label{fig:optical_gain}
		The imaginary part of the effective refractive index for the two modes propagating along the $[1 1 0]$ direction of the photo-excited GaN crystal. 
		The insets display the polarization directions ($\mathbf{e}_{1}$ and $\mathbf{e}_{2}$) relative to the crystal orientation.
		The blue and red circles represent the positions of nitrogen and gallium atoms, respectively. 		
	}
\end{figure}

With the aid of $\chi_{\alpha \beta}(\omega)$, we investigate the linear propagation of light in the direction that is perpendicular to both the polarization direction of the excitation pulse and the crystal axis.
Let $\mathbf{u}$ be a unit vector pointing in this direction, the Miller indices of which are $[1 1 0]$.
The two waves that propagate in this direction without changing their polarization state are found by solving the following eigenproblem:
\begin{equation}\label{eq:eigenmodes}
\hat{\epsilon}^{-1}(\omega) \left(\mathbb{1}-\mathbf{u} \otimes \mathbf{u}^{\mathrm{T}} \right) \mathbf{e}_{i}(\omega) = n_{i}^{-2}(\omega) \mathbf{e}_{i}(\omega),
\end{equation}
where $\hat{\epsilon}(\omega) = \mathbb{1}+ 4\pi \hat{\chi}(\omega) $ is the permittivity tensor.
Doing so, we discard the solution with zero eigenvalue, where $\mathbf{e}_{i} \parallel \mathbf{u}$.
The other two eigenvectors, $\mathbf{e}_{1}$ and $\mathbf{e}_{2}$, are the polarization vectors of the two propagating modes.
Since we chose $\mathbf{e}_L$ to be in the mirror plane of the crystal, photo-excitation by the laser pulse preserves the mirror symmetry.
Because of this, the two modes are parallel to either $\mathbf{e}_L$ or the crystal axis.
The $n_{i}^{-2}$ eigenvalues in Eq.~\eqref{eq:eigenmodes} are inverse squares of the effective refractive indices for both modes.
A mode experiences optical gain if the imaginary part of its refractive index is negative.
For the mode that is polarized along $\mathbf{e}_L$, we plot $\text{Im}[n]$ in Fig.~\ref{fig:optical_gain}(a).
This mode mainly experiences absorption.
The other mode, which is polarized along the crystal axis, is strongly amplified, as we see in Fig.~\ref{fig:optical_gain}(b).

Comparing Fig.~\ref{fig:population_inversion} and Fig.~\ref{fig:optical_gain}, we  see that the $\mathbf{e}_{1} \parallel \mathbf{e}_L$ mode is particularly efficient at driving transitions that absorb light, while the other mode benefits most from population inversion.
This difference can be understood by analyzing the matrix elements that are responsible for transitions that shaped Fig.~\ref{fig:population_inversion}.
For the star of $\mathbf{k}$ that is responsible for optical gain at 0.6~eV, $\max_{\mathbf{k}} \abs{\mathbf{e}_{2} \cdot \mathbf{d}_{\mathrm{SO-LH}}(\mathbf{k}) }$ exceeds $\max_{\mathbf{k}} \abs{\mathbf{e}_{1} \cdot \mathbf{d}_{\mathrm{SO-LH}}(\mathbf{k}) }$ by a factor of 15; for the 1.1-eV star of $\mathbf{k}$, this is a factor of 6.
Consequently, the $\mathbf{e}_{2}$ mode more efficiently drives the transitions that are responsible for stimulated emission.
In other words, the $\mathbf{d}_{\mathrm{SO-LH}}(\mathbf{k})$ matrix elements for these transitions are approximately aligned with the crystal axis, so that they do not amplify the $\mathbf{e}_{1}$ mode, which experiences absorption by driving transitions between other pairs of valence- and conduction-band states, where no population inversion is present.

\begin{figure}
	\includegraphics[width=0.92\columnwidth]{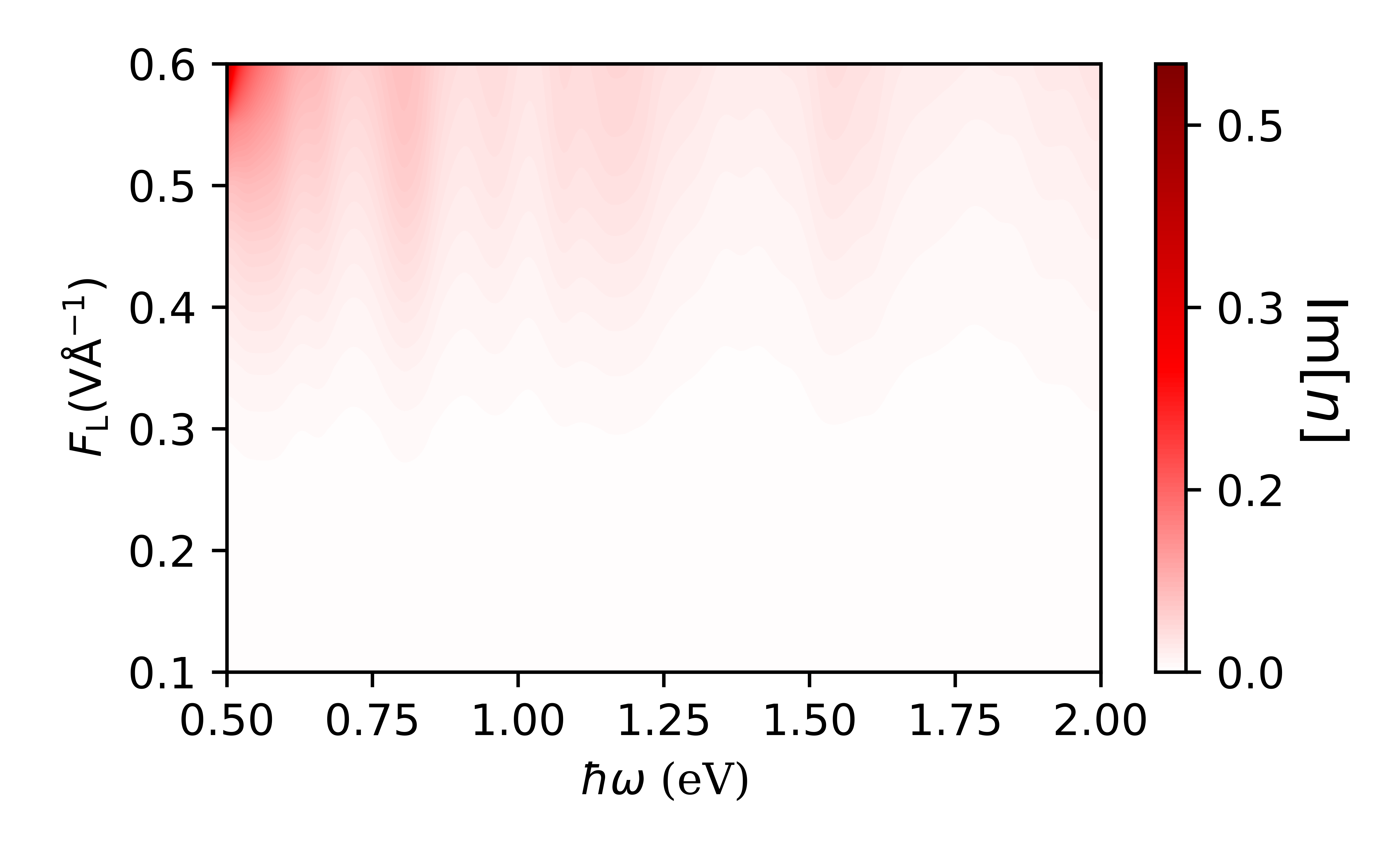}
	\vspace{-5mm}
	\caption{\label{fig:thermal_gain}
		The imaginary part of the refractive index after thermalization, assuming a weak probe light polarized along the crystal axis and propagating in the $[1 1 0]$ direction.
	}
\end{figure}

Photo-excited charge carriers undergo fast relaxation dynamics by interacting among themselves, as well as with phonons.
While the charge-carrier recombination is a relatively slow (nanosecond-scale) process, it takes merely a few tens of femtoseconds for the energy distributions of electrons and holes to approach those prescribed by the Fermi-Dirac statistics~\cite{Shah_1999, Rohde_PRL_2018}.
In this thermalized state, occupations monotonously decrease with increasing state energy, so population inversion is quickly destroyed by relaxation.
While the theoretical description of femtosecond-scale relaxation is challenging, it is easy to obtain state occupations in a thermalized state assuming that the total energy and the concentration of charge carriers are preserved during thermalization~\cite{Sato_PRB_2014_temperature}.
By applying this procedure to $f_n(\mathbf{k}, F_L)$, we obtain occupation probabilities in the thermalized state, $f_n^{\mathrm{thermal}}(\mathbf{k}, F_L)$.
Then we evaluate the optical response of these thermalized states using Eqs.~\eqref{eq:chi} and \eqref{eq:eigenmodes}.
Thermalization does not change the polarization states of the two modes propagating in the $[1 1 0]$ direction, but it profoundly changes the optical properties of the photoexcited crystal, as we show  in Fig.~\ref{fig:thermal_gain}: there is no optical gain, the peak values $\text{Im}[n]$ are much smaller than those immediately after the photoexciting laser pulse, but the absorption is more homogeneously distributed over the displayed spectral range.
These results are similar for both modes, so Fig.~\ref{fig:thermal_gain} shows $\text{Im}[n(\omega, F_L)]$ only for $\mathbf{e}_2$---the mode that experienced strong optical gain before thermalization.

In summary, we have described conditions that are particularly favorable for inducing population inversion and optical gain by ultrafast, strong-field photoinjection in a transparent uniaxial crystal.
We have also provided numerical evidence that, for carefully chosen directions of the crystal axis, pump and probe fields, this is expected to be a significant effect (propagation over a distance as short as a single wavelength doubles light intensity for $\text{Im}[n]= -0.055$).
In reality, the amplification must be weaker because, even with few-cycle pump and probe pulses, relaxation will reduce population inversion during the light-matter interaction---Fig.~\ref{fig:optical_gain}(b) presents an an upper limit for small-signal gain.
Nevertheless, as long as population inversion formed between valence bands persists for $\gtrsim 10$~fs, this should be a measurable effect.
As a potential application, the short-lived optical gain may assist purely optical investigations of the fastest stages of relaxation that electrons and holes undergo after ultrafast photoinjection by an intense laser pulse.
Currently, little is known about the relaxation of such very nonequilibrium electronic excitations, especially in the presence of a strong laser field.
In this context, field-resolved measurements may become particularly important---it is nowadays possible to measure how transmission through a thin sample changes the time-dependent electric field of a light wave~\cite{Sommer_Nature_2016}.
In this case, the temporal resolution of a pump-probe measurement is not limited by the duration of the probe pulse, which gives access to the processes unfolding within a fraction of an optical cycle.
Transient optical gain is well suited for such investigations because it unambiguously indicates the presence of population inversion.

\begin{acknowledgments}
	The authors are grateful to Ferenc Krausz for his support and stimulating discussions, to Michael Wismer for developing the code, and to Matthew Weidman for commenting on the manuscript.
	M.\,Q.\ and D.\,A.\,Z.\ were supported by the International Max Planck Research School of Advanced Photon Science (IMPRS-APS).
\end{acknowledgments}

\end{document}